\documentstyle[aps,prb,twocolumn,epsfig]{revtex}
\begin{document}
\def\be{\begin{equation}}
\def\ee{\end{equation}}

\title{Dimension dependence of the conductance distribution in the non-metallic regimes}

\author{Peter Marko\v s$^*$\\
  Institute of Physics, Slovak Academy of Sciences,
  D\'ubravsk\'a  cesta 9, 842 28 Bratislava, Slovakia}
\maketitle                                 

\begin{abstract}

The conductance distribution $P(g)$ in disordered three-dimensional systems
in  the critical and insulating  regimes is calculated numerically and compared with
$P(g)$ of  weakly disordered quasi-one dimensional systems.
Although both system  exhibit similar transport properties,
their conductance distributions differ quantitatively.
We explain this difference by analysis of the  spectrum of transfer matrix. 
Our numerical data confirm  that the statistics of the conductance
 is consistent with the one-parameter scaling also in the strongly localized
regime.
\end{abstract}

\section{Introduction}

Our understanding of the 
metal - insulator transition (MIT) in disordered solids is based
on the scaling theory of localization 
\cite{AALR}. We believe that MIT is universal. It means that the 
system-size dependence of any quantity, say conductance $g$, is 
determined only by one parameter.
The validity of the one parameter scaling was tested and confirmed 
in the neighborhood of the critical point  by
various  numerical simulations on quasi-one dimensional (Q1D) systems
\cite{MacKK,q1d}.  However,  a detailed analytical theory
of MIT is still missing. Their formulation is complicated also by 
absence of the self-averaging of the conductance in disordered systems
at zero temperature
\cite{shapiro}. Complete description of transport properties 
requires  therefore 
a detailed analysis of the  conductance distribution $P(g)$. Such 
theory  exists only 
in the  weak disorder limit. The weak disorder expansion\cite{stone},   
Dorokhov-Mello-Pereyra-Kumar equation (DMPK)
\cite{DMPK},
and the random matrix theory (RMT) 
\cite{pichard,beenak}
provide us with entire quantitative description of the transport statistics.
However, critical regime never appears in the weak disorder limit. 

For  the critical regime,
analytical results for the conductance statistics 
were derived only in dimensions slightly above the lower critical dimension
$d=2+\varepsilon$  with $\varepsilon\ll 1$
\cite{epsilon,epstwo}.
These cannot be extrapolated to three dimensional systems ($\varepsilon=1$)
\cite{MKa}. 
Our information about the critical regime in 3D systems 
is  therefore  based only on  numerical  simulations.
The shape  of the conductance distribution at the critical point 
was analyzed
\cite{MKa,ando,SO,PRL,RMS,soukou} 
and one-parameter scaling of the mean conductance was numerically proven in 
\cite{SMO}. 

Recently, an attempt has been made to obtain relevant information about 
the critical regime from studies of the weakly disordered Q1D systems
\cite{MW,MWa}.
Although Q1D systems  do not exhibit critical phenomena 
(correlation length never diverges )
 their transport properties resemble 
that observed in 3D systems: (1) 
in the metallic regime
when the system length $L_z$ is shorter than the localization length $\xi$,
both RMT and  the DMPK equation predict that $P(g)$ in Q1D systems is 
Gaussian with a constant, disorder independent width. 
This was numerically  observed also 
for squares (2D) \cite{pozn}  and cubes (3D) 
\cite{pichard,ostatni,MKa}. Only the variance of conductance  
depends on the dimension
\cite{stone}. 
(2) 
In the limit of very long system length, $L_z\gg\xi$, 
Q1D samples become insulating,
in spite of weak disorder.  RMT predicts  that
the distribution of $\log g$ 
is  Gaussian and var  log $g = -2 \langle\log g\rangle$
\cite{pichard}. Log-normal distribution was observed also in 2D and 3D 
strongly disordered systems
\cite{MKb,MKc}. The dimension-dependence of the variance var log $g$
is, however, not understood.
(3) In the intermediate regime, where 
$\langle g\rangle\sim 1$, analytical relation for 
the   conductance distribution can be obtained
\cite{MW,MWa}.  Its  shape is  similar
to that obtained numerically at the critical point of MIT
\cite{RMS}. 

\medskip

In this paper we discuss the differences between 
the   conductance distribution for Q1D weakly disordered systems
and for 3D cubes. As the metallic limit is already good understood
both analytically
\cite{DMPK,pichard,beenak,imry}
and numerically
\cite{MKa,ostatni}
we concentrate to the critical regime ($\langle g\rangle\sim 1$)
and the strongly localized regime ($\langle g\rangle<<1$.
Although the distributions are qualitatively
similar in each transport regime, quantitative differences will be found.
We show that statistics of the conductance depends on the dimension in the
critical and even in the localized regime. This must be taken into account
in  the entire description of the transport. Our analysis of  the 
statistical properties 
of parameters $z$, which are defined by Eq. (\ref{g}) below, 
confirms the validity
of the one-parameter scaling theory  in the insulating regime.

In  Section 2 we describe the  model we use. 
Section 3 presents 
numerical data for the conductance  distribution in the intermediate
(critical) regime.  
The shape of the distribution in the three dimensional 
localized regime is studied in detail in Section 4. 

\section{The model}

We consider  Anderson model defined by the   Hamiltonian
\begin{equation}\label{anderson}
{\cal H}=W\sum_n\varepsilon_n|n\rangle\langle n|+\sum_{[nn']}
|n\rangle\langle n'|+ |n'\rangle\langle n|
\end{equation}
where $n$ counts the  sites in $d$-dimensional lattice  and $nn'$ are nearest-neighbor sites. The size of the lattice is either $L^d$ ($d=2,3$) 
or $L^{d-1}\times L_z$. 
Random energies $\varepsilon_n$ are distributed  with the box distribution 
$|\varepsilon_n|\le 1/2$ and $W$ is the strength
of the disorder.

Our analysis of the conductance distribution
is based on the Landauer formula for the conductance \cite{landauer}
\be\label{g}
g= {\rm Tr}~~ t^\dag t= \sum_1^N\cosh^{-2}(z_i/2)
\ee
In (\ref{g}), 
$t$ is the transmission matrix, and  the parameters $z_i$,
$i=1,2,\dots,N$, the parameterize the eigenvalues   of $t^\dag t$
 ~\cite{pichard}.
In the limit $L_z>>L$, parameters $z$ converge to Lyapunov exponents of the
transfer matrix.
$N$ is the number of open channels 
\cite{aando}.

According to (\ref{g}), 
statistics of the conductance is completely determined 
by statistical properties of the parameters $z$. 
It is well known that
in  weakly disordered systems the spectrum of $z$ is linear
\be\label{linear}
z_i\propto i
\ee
independently
on  dimension and shape of the sample\cite{DMPK,pichard}.
The linear behavior (\ref{linear}) 
can be  derived also from the random matrix model
\cite{muttalib} and  was used  in  Ref. \cite{imry} to
prove of the universality of 
conductance fluctuation \cite{stone}.

It is important to mention  that weakness of the  disorder is  
not only necessary but also sufficient 
condition for the linear form of the spectra (\ref{linear}). 
Relation (\ref{linear}) holds therefore for any weakly disordered Q1D systems 
independently on the system   length
\cite{MacKK,pichard,beenak,muttalib}.
This is not true when disorder increases.
Thus, we have numerically proven that
\be\label{kvad}
z_i^{d-1}\propto i \quad\quad W=W_c.
\ee
at the critical point of MIT in dimension $d=3$ and 4.
Relation (\ref{kvad})  holds for  both Q1D \cite{nato}  and 3D \cite{MKa} systems.
In  3D, it enables us to understand  the dimension-dependence of the
conductance distribution at the critical point 
\cite{PRL,RMS}. 

The form of the spectra of $z$  for strongly disordered systems is known
for  the Q1D samples \cite{JPCM}. 
An attempt was made to study $z$ for 2D and 3D strongly disordered 
systems \cite{MKb,MKc}. We present here much more accurate data for parameters 
$z$, which
enable us to estimate  the contribution of higher channels to the conductance.

\section{Conductance distribution in the intermediate (critical) regime}

Three-dimensional  Anderson model with a box distributed disorder exhibits MIT
at the critical point $W_c=16.5$. 
We calculated the conductance distribution
for ensembles of three-dimensional cubes with disorder 
$10.6\le W\le 21.5$. The system size was  $8\times 8\times 8$ which is  
large enough to exhibit quantitative properties of the conductance distribution.
Calculated shape of $P(g)$ 
is  compared with the conductance distribution for weakly disordered
Q1D systems of the size $8^2\times L_z$. 
The disorder $W=4$ assures the existence of the metallic regime 
for short $L_z$, $L_z\le 100$.
The localized regime appears when $L_z$ exceeds 250.
Tuning the length $L_z$ we find for each 3D ensemble the 
corresponding Q1D ensemble  with the same mean conductance.

\begin{figure}[t]
\noindent\epsfig{file=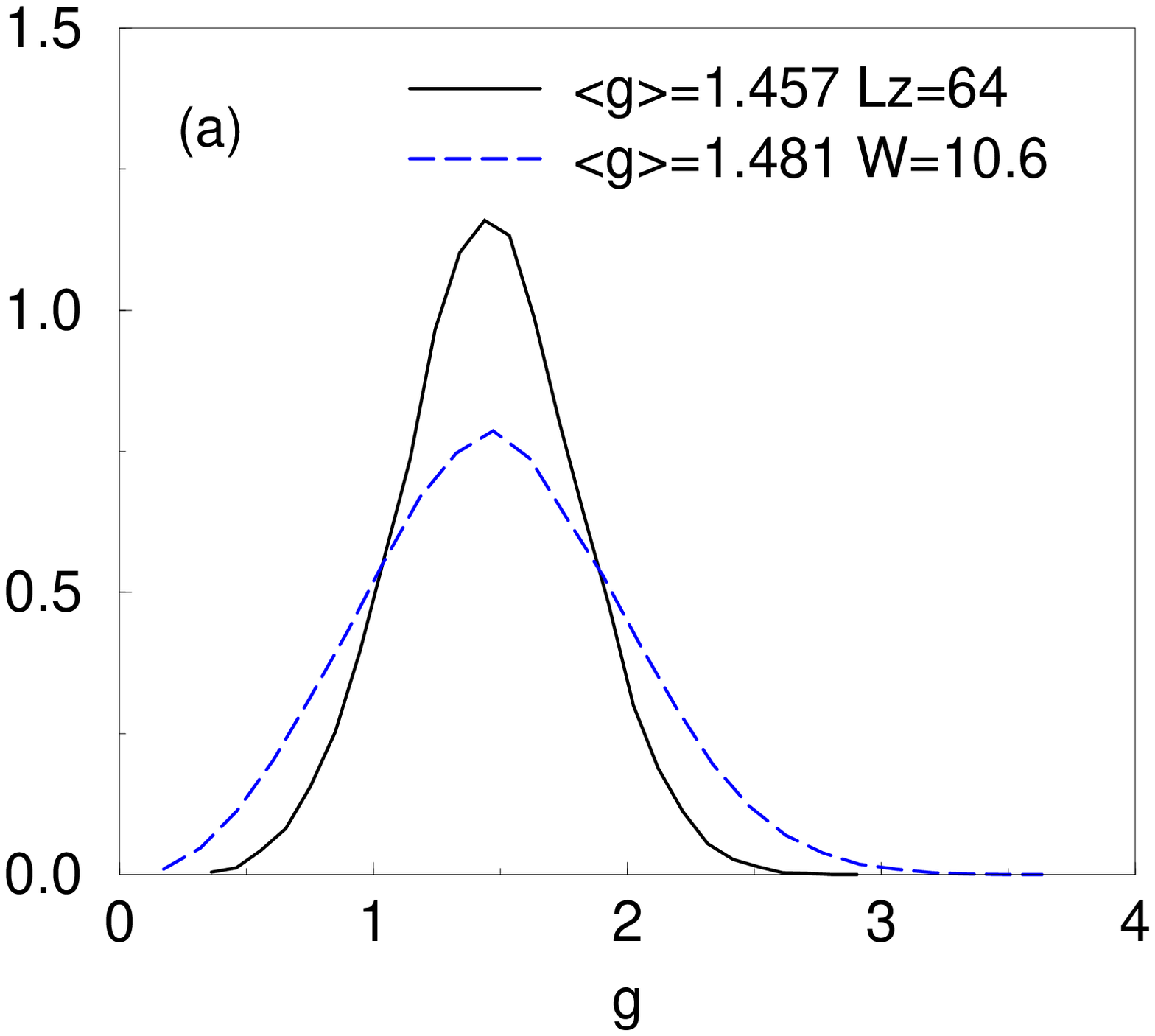,width=4cm}~~\epsfig{file=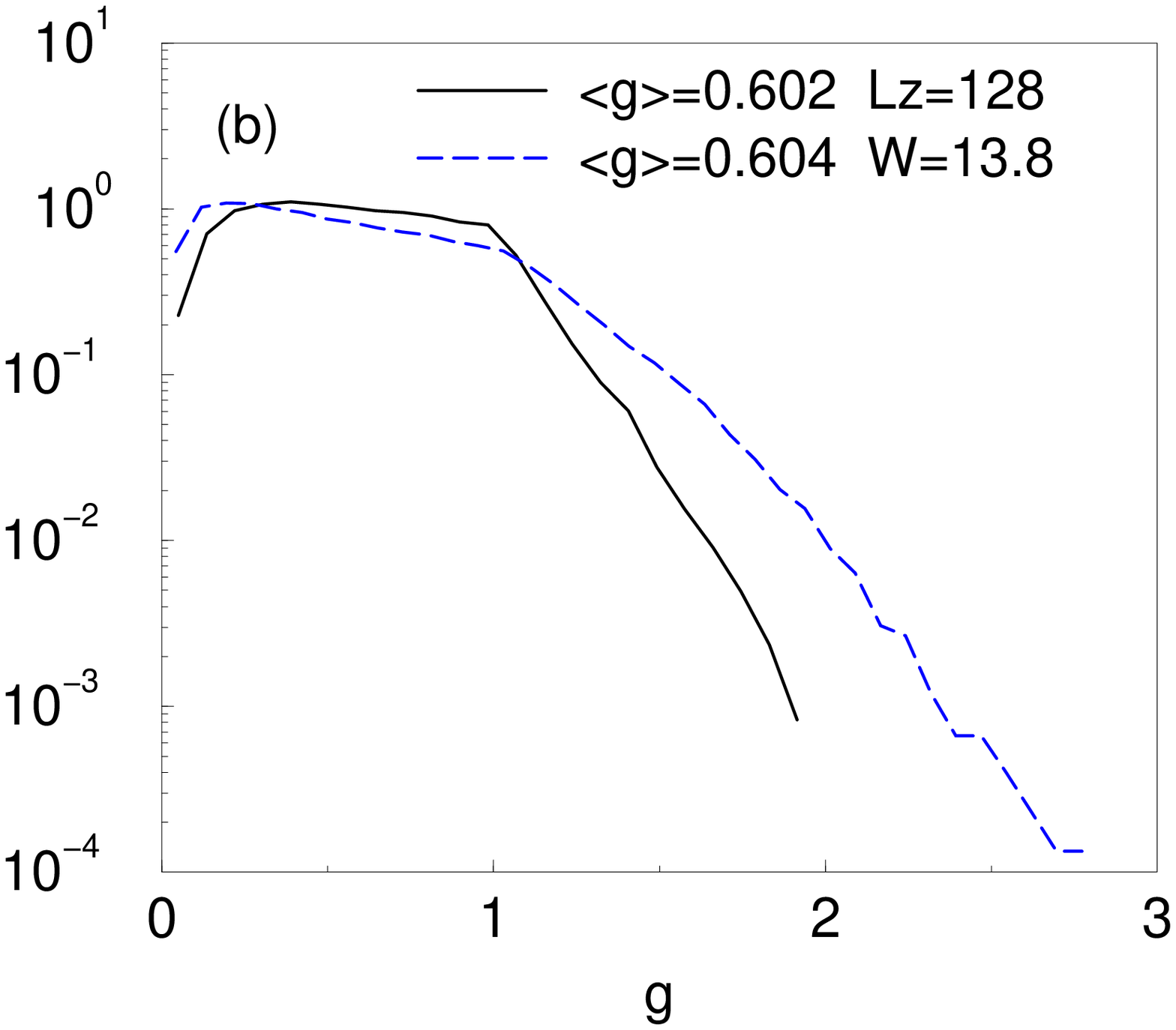,width=4cm}
\noindent\epsfig{file=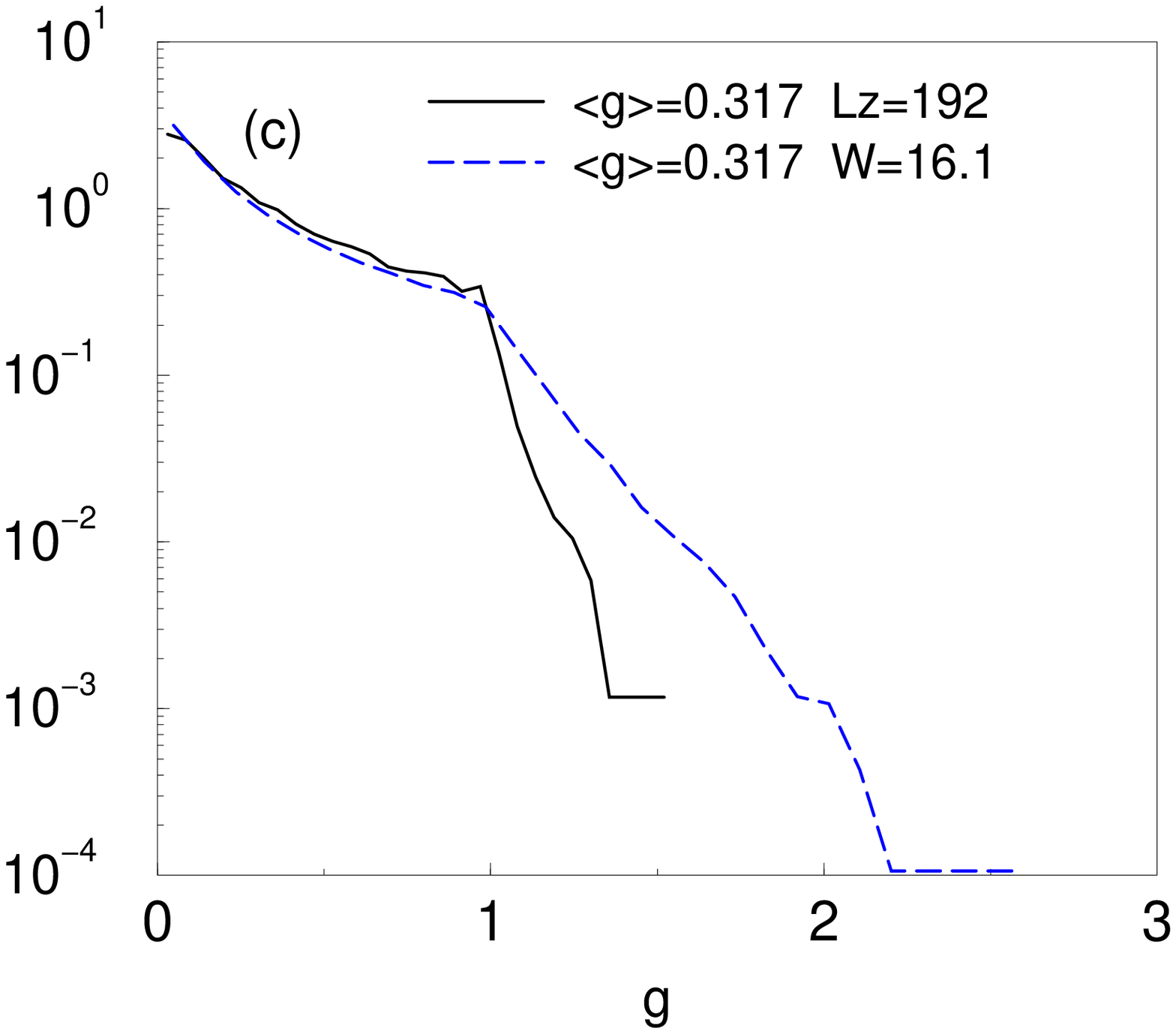,width=4cm}~~\epsfig{file=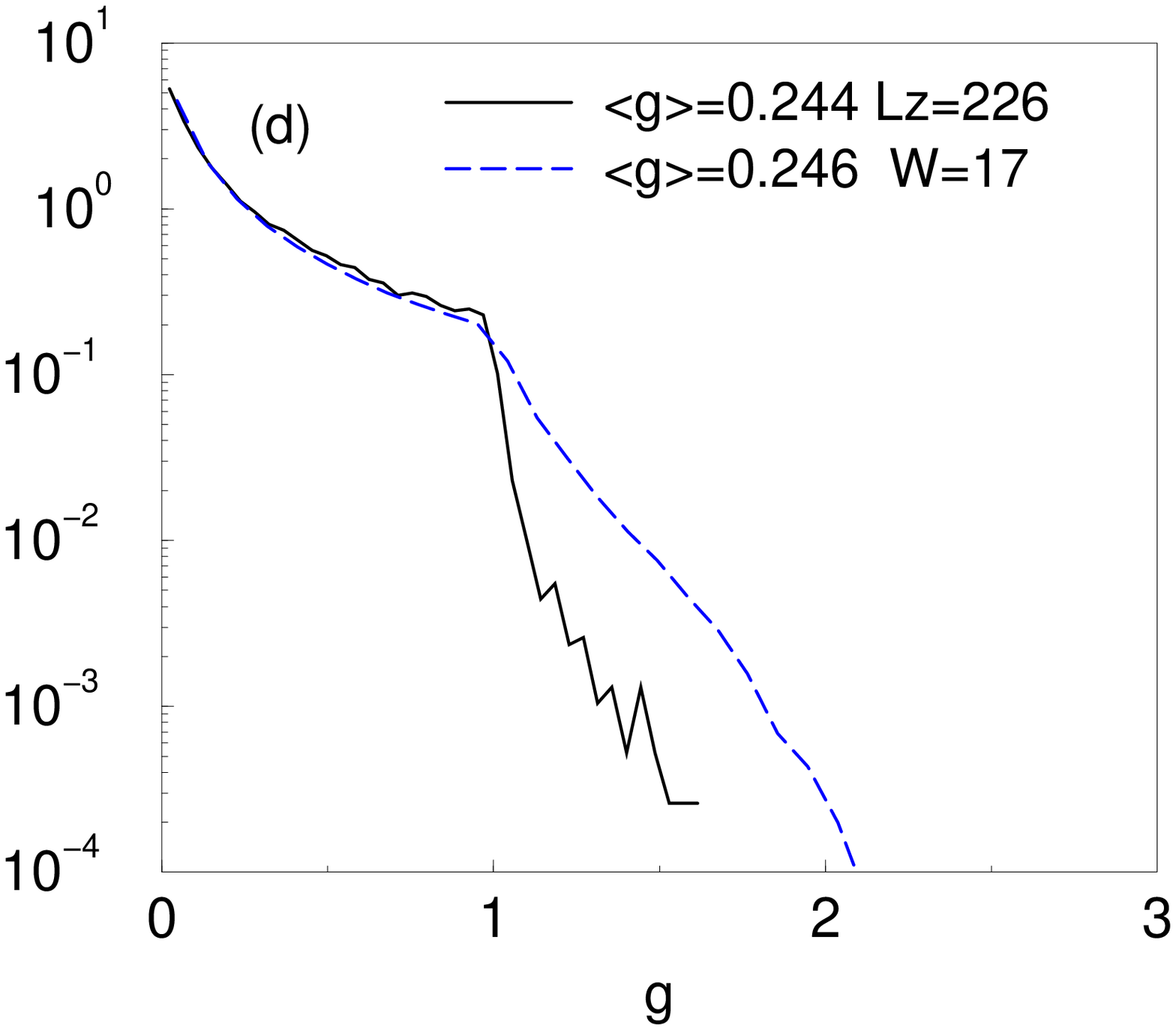,width=4cm}
\noindent\epsfig{file=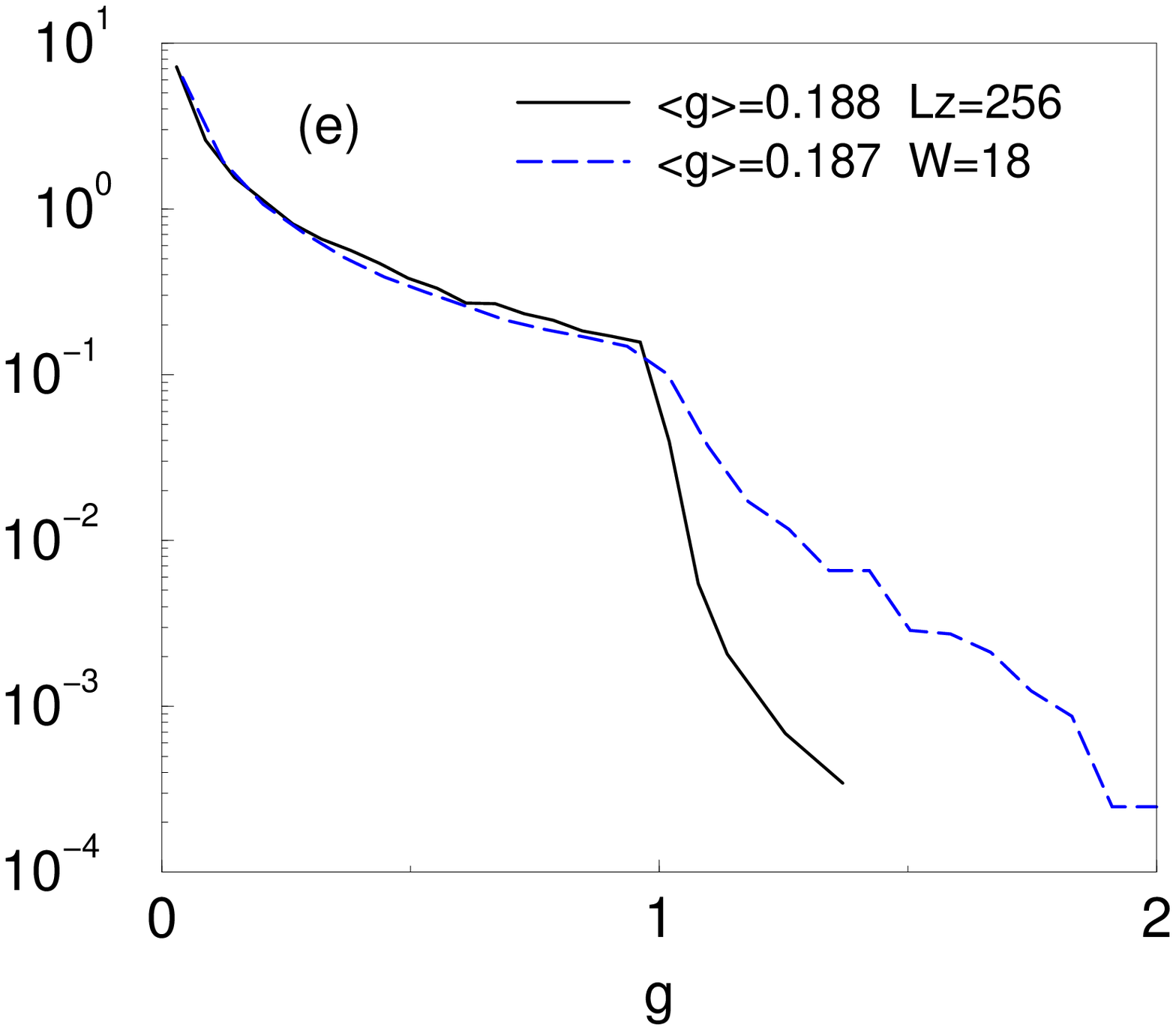,width=4cm}~~\epsfig{file=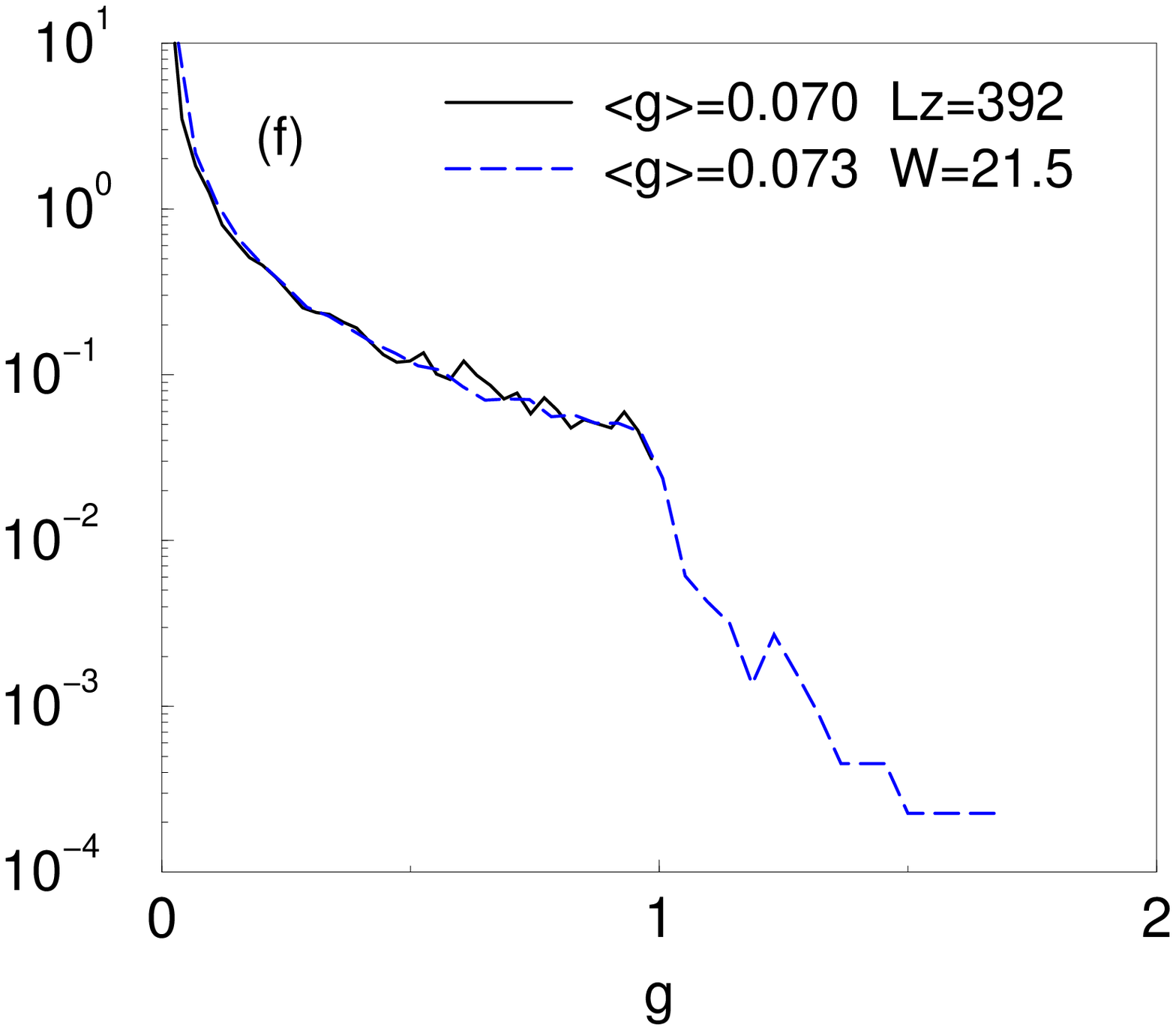,width=4cm}
\vspace*{2mm}
\caption{Conductance distribution $P(g)$ for  
cubic samples  $L\times L\times L$ with $L=8$. 
Disorder increases from $W=10.6$ (metallic regime, $\langle g\rangle=1.481$) to $W=21.5$
(insulator, $\langle g\rangle=0.073$).
Critical disorder is $W_c=16.5$ and mean  
conductance at the critical point is $\langle g\rangle_c\approx 0.28$.
For each value of the disorder, we plot $P(g)$ (dashed line) and compare it
with the conductance distribution for weakly disordered Q1D system 
which posses the same  mean conductance (solid line). 
Size of Q1D system  is $L^2\times L_z$ with  $L=8$, disorder $W=4$ and length
is tuned to get required $\langle g\rangle$.
Fixed boundary conditions were applied in $x$ and $y$ directions.
$10^5$ samples were collected for most statistical ensembles.
(a) - metallic regime, $P(g)$ is Gaussian for both cubic and Q1D geometry.
Distributions differ from each other only in their width:
var $g_{\rm 3D}\approx 0.27$ and var $g_{\rm Q1D}\approx 0.12$. This agrees with
theoretical data (0.31 and 0.133 for 3D and Q1D, respectively\cite{stone}).
Figures (b) - (e)  present $P(g)$ in the 
critical regime: $\langle g\rangle\le 1$. The main difference between 
conductance distributions is observable in region $g>1$.
(f) shows the conductance  distributions
in the insulating regime. Note that there are no Q1D sample with conductance $g>1$
while $P(g)$ for cubes still possesses tail up to $g\approx 1.5$
}
\label{fi1}
\end{figure}

\begin{figure}
\centerline{\resizebox{8.25cm}{6.25cm}{\includegraphics{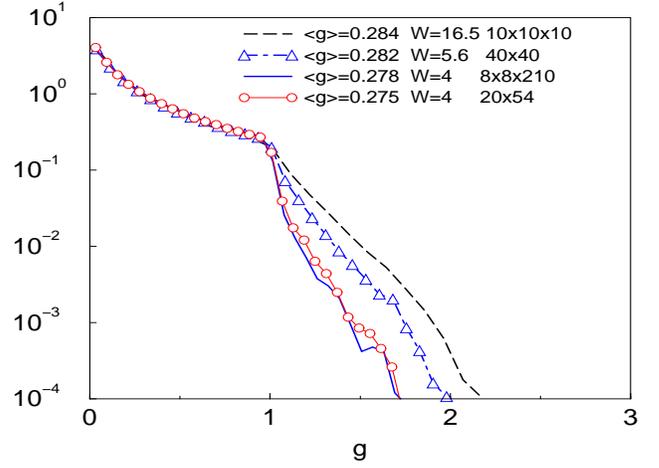}}}
\caption{Critical conductance distribution for 3D Anderson model
(dashed line) and conductance distribution for square and quasi-one
dimensional systems $L\times L_z$ and $L^2\times L_z$ which have
approximately the same mean conductance. As it is supposed,
the quasi-one dimensional systems have the identical distributions.
}\label{fi1a}
\end{figure}

A comparison of the conductance distribution 
of 3D and Q1D systems  is presented in Figures \ref{fi1}, \ref{fi1a}
and \ref{fi3}.
In the metallic regime, 
when $\langle g\rangle>1$, $P(g)$
is Gaussian in both 3D and Q1D geometry, as supposed
(Fig. \ref{fi1}a).
The width of the distribution is universal, independent on the system size. 
It depends only on the
system shape, and is slightly smaller for Q1D than for 3D systems, in
agreement with theoretical results
\cite{stone,imry}.

Figures 1 (b-e) present the shape of $P(g)$ 
in the critical regime, when  $\langle g\rangle\approx 1$. We know that 
$P(g)$  for 3D systems becomes non-analytical at $g=1$ and 
decreases exponentially when  $g>1$ 
\cite{PRL,RMS}.
The same  is true also for Q1D systems in the intermediate regime
\cite{MW,MWa}. $P(g)$ decreases for $g>1$ much faster than predicted
theoretically
\cite{MW}. 
The exponential  decrease in Q1D is even 
faster than that in 3D systems.
This is because 
 for a given $\langle z_1\rangle$ the 
difference $\langle z_2-z_1\rangle$ 
is much smaller in 3D than in Q1D, as follows from the comparison of 
(\ref{linear}) with   (\ref{kvad}).
Higher channels have therefore a 
better chance to contribute  to the conductance in 3D than in Q1D. 

Figure \ref{fi1a} presents the conductance distribution for 
3D Anderson model at the critical point ($W_c=16.5$) and for 
2D and Q1D systems with the same mean conductance. As supposed,
3D distribution posses the longest tail, while two Q1D-distributions
are almost identical. From the  plot of the $P(\log g)$ (data not shown in 
the paper) we conclude that  all four distributions are identical for small
$g$. This is not surprise since the small -$g$ behavior is determined
completely by the distribution of $z_1$, which is the same in all systems.

\begin{figure}[t]
\centerline{\resizebox{8.25cm}{6.25cm}{\includegraphics{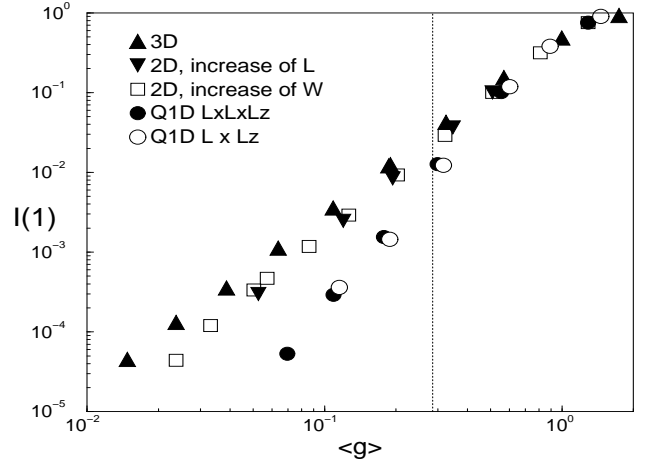}}}
\vspace*{2mm}
\caption{Integral $I(1)=\int_1^\infty P(g){\rm d}g$  as a function of $\langle g\rangle$ for different geometrical shapes of samples. 
Dotted vertical line indicates mean critical conductance for 3D Anderson model.
$I(1)$ decreases  more quickly for Q1D systems than for squares and cubes.
For the quasi-one dimensional systems data 
for systems $L\times L_z$ and $L^2\times L_z$ scale to one curve. 
This is caused by  of the universal (linear) spectrum of $z$. 
Data for squares and cubes  scale differently due to the
dimension dependence of spectrum of $z$s.
}\label{fi2}
\end{figure}

\begin{figure}
\noindent\epsfig{file=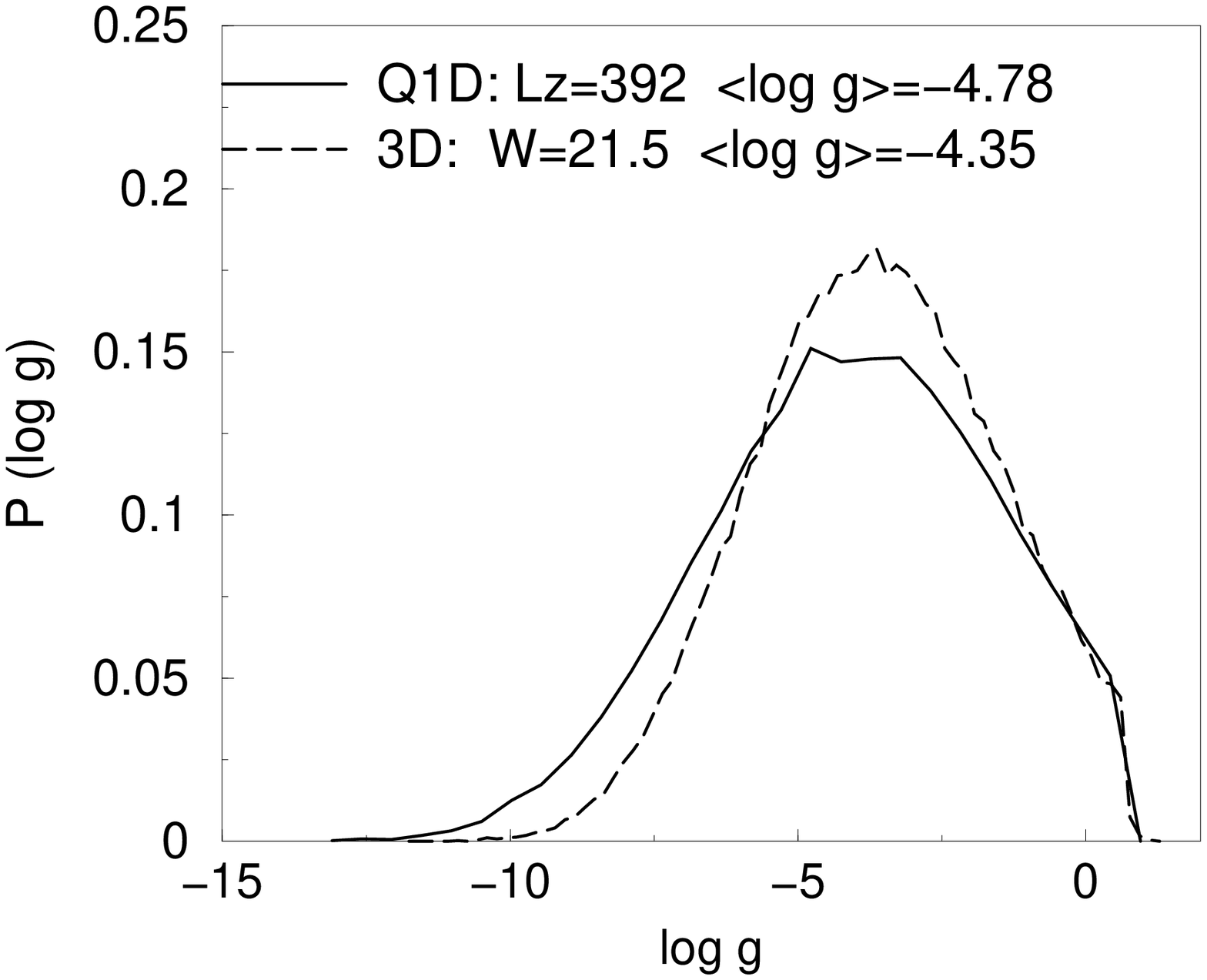,width=4cm}~~\epsfig{file=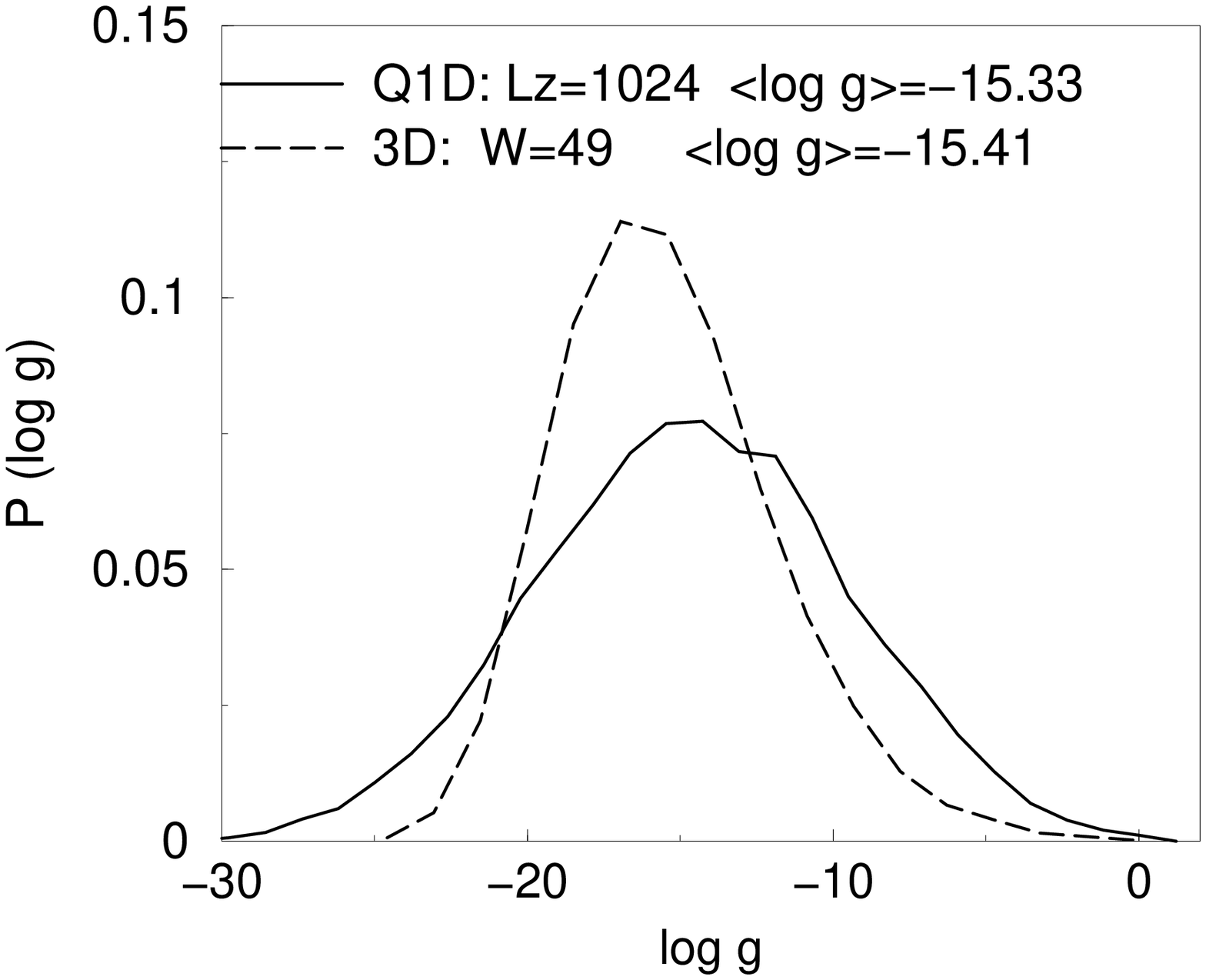,width=4cm}
\caption{(a) Distribution $P(\log g)$ for 3D and Q1D systems  
displayed in Figure \ref{fi1} (f). Note that although systems have the same
$\langle g\rangle$, they differ in $\langle\log g\rangle$.
This indicates that mean conductance is not relevant parameter in the
insulating regime. The width of the distribution is larger for Q1D system,
(var log $g\approx 6$ for the Q1D system, and $\approx 4.33$ for 3D system). 
(b) the same for 3D and Q1D  systems with $\langle\log g\rangle\approx -15.3$. Different width of distribution is clearly visible:
var log $g\approx 25.66$ for the Q1D system, and only $\approx 12.6$ for 3D system. 
This is consistent with previous numerical results
\cite{pichard}. 
}\label{fi3}
\end{figure}

The quantitative difference between the conductance distribution for
cubic and Q1D systems is measured by the integral
\be
I(1)=\int_1^\infty P(g) {\rm d}g
\ee
In figure \ref{fi2} we plot $I(1)$ as a function of 
the mean conductance $\langle g\rangle$ for various
geometrical shapes of  systems. 
For Q1D systems, we calculated both $L\times L_z$ and $L^2\times 
L_z$ ($L_z>>L$).  The data  lay  on  the same 
curve.  This is no surprise  since the spectrum of $z$ is linear in
both systems.
A similar universality was found in 2D system. 
Here, we either increase disorder with constant system size
or increase the size of square samples keeping disorder constant.
As  supposed, the data for 3D and 2D systems scales independently.

The figure \ref{fi3} compares  $P(\log g)$ for 3D and Q1D systems 
in localized regime.  It confirms that 
$P(\log g)$ converges to Gaussian  both for 3D and Q1D systems. 
The distribution
for Q1D systems is  much broader than that for 3D.
While the shape of $P(\log g)$ in weakly disordered
Q1D systems can be obtained analytically
\cite{pichard,muttalib}, distribution for 3D systems 
is known only from numerical simulations
\cite{MKb,MKc} 
and deserves  more detailed analysis.

\section{Strongly localized regime}

In the insulating regime  all $z$ increase linearly with the system length.
The length-dependence of $z_1$ determines the localization length $\xi$
\be\label{locl}
z_1(L)=\frac{2L}{\xi}+{\rm const}
\ee
(For Q1D systems, $L$ should be replaced by $L_z$).

\begin{figure}
\centerline{\epsfig{file=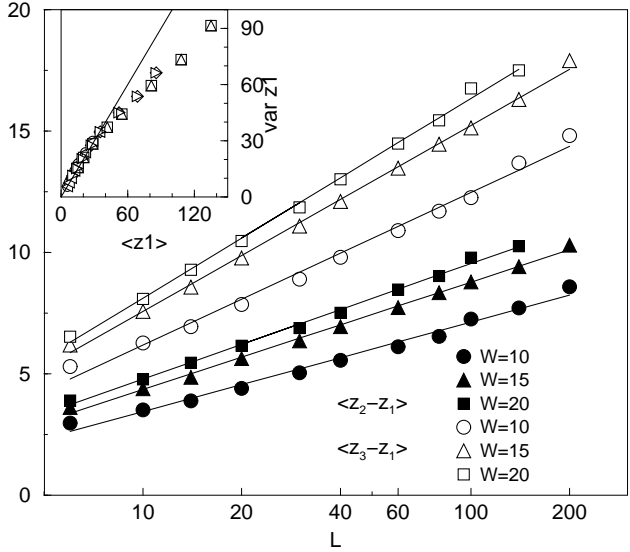,width=8.25cm}}
\bigskip
\caption{System-size dependence of the differences $\langle z_2-z_1\rangle$ 
and $\langle z_3-z_1\rangle$
for 2D orthogonal systems. Three different values of disorder were used. 
Solid lines are logarithmic
fits. 2D symplectic systems exhibit the same logarithmic
dependence.
Inset: Plot of var $z_1$ vs. $\langle z_1\rangle$.}\label{fi4}
\end{figure}

The knowledge of the spectrum of parameters 
$z$ is important also in the insulating regime.
As mentioned  above, the spectrum of $z$ remains linear 
in weakly disordered Q1D
systems \cite{pichard,MacKK,muttalib}
so  that $z_2=2 z_1\gg z_1$.
Contrary   to Q1D systems, we find that 
the difference $\langle z_2-z_1\rangle$ increases only logarithmically with the system
size for 2D systems (Figure \ref{fi4}):
\be\label{log}
\langle z_2-z_1\rangle\sim c\times\log L\quad\quad {\rm 2D~systems}.
\ee 
The same logarithmic behavior 
was found also in the 2D systems with spin-orbit scattering (data are not presented here).

\begin{figure}[t]
\centerline{\epsfig{file=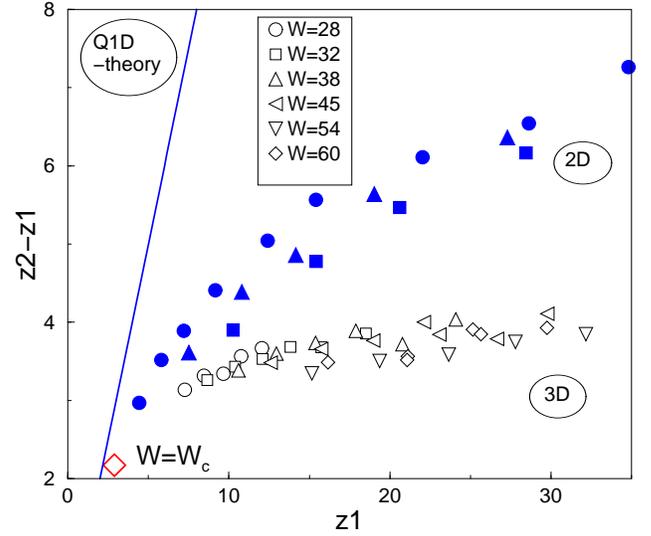,width=8.25cm}}
\caption{$\langle z_1\rangle$-dependence of  difference $\langle z_2-z_1\rangle$
for 2D (full symbols) and 3D (open symbols) disordered systems.
Solid line is the theoretical prediction 
$\langle z_2-z_1\rangle=\langle z_1\rangle$ for the Q1D weakly disordered
systems.
}\label{fi5}
\end{figure}

In 3D systems we find that 
the differences 
$\langle z_2-z_1\rangle$ and
$\langle z_3-z_1\rangle\dots$ do not
depend on the system size  in the strongly localized regime:
\be\label{threed}
\langle z_2- z_1\rangle\approx {\rm const}\quad\quad{\rm 3D~systems}
\ee
In figure \ref{fi5} we present the $z_1$-dependence of the
 difference $\langle z_2-z_1\rangle$ for the 2D and 3D systems.
For $z_1$ large enough, data scale to the same curve, confirming the
one-parameter scaling. More important, data prove that the spectrum of
the transfer matrix depends on the dimension of the sample in the
localized regime. 

\begin{figure}[t]
\centerline{\resizebox{8.25cm}{6.25cm}{\includegraphics{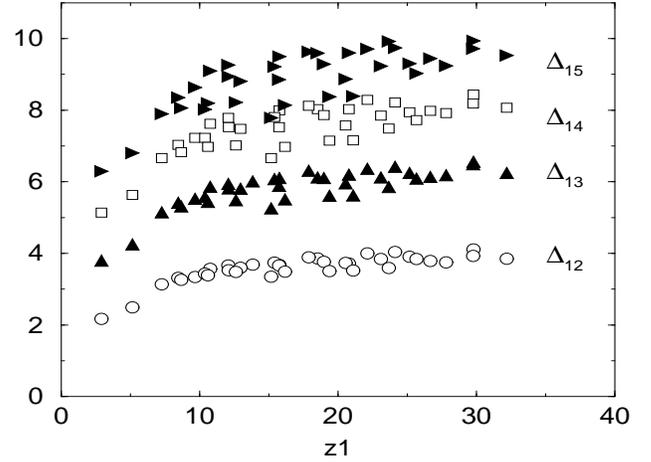}}}
\vspace*{3mm}
\caption{%
$z_1$-dependence of the differences $\Delta_{i1}\langle z_i-z_1\rangle$ for
$i=2,3,4$ and 5.  Disorder $W=$ 28, 32, 38, 45, 54 and 60,
system size $L=6,8,\dots 16$. Data for $W=16.5$ and $W=20$ ($L=8$) are also
presented to show how $\Delta$s 
$\Delta$s increase  from their "critical" values ($W=W_c$). For large $z_1$,
they  converges to constant, which do not 
depend neither on the system size nor on the disorder.
}\label{fi6}
\end{figure}

Figure \ref{fi6} presents the $z_1$ - dependence of the differences
$\langle z_i-z_1\rangle$ for $i=2,3,4$ and 5. Although the convergence 
is worse for higher differences, the saturation to the $z_1$-independent
constant is clearly observable as $z_1$ increases.

Due to (\ref{threed}), the contribution of the second channel is nonzero
whichever is the contribution of the first one.
This
contribution, however gives   only small correction to  
the first channel contribution $\langle \log g\rangle=\log 4 -\langle z_1\rangle$. Nevertheless, 
it is interesting to test how  the contributions of the higher
channels influence the statistical properties of the conductance.
We found that apart   the  shift of the mean value
the distribution $P(-\log g)$ and the distribution $P(z_1)$ 
are almost identical.

Although the absolute value of the conductance changes only slightly
due to the contribution of higher channels, the present effect 
influences  the variance of the log $g$. Indeed, the small difference
$z_2-z_1$ causes that 
$z_1$ is confined from above: the probability to find  
$z_1>>\langle z_1\rangle$
is much less in the 3D than in the Q1D.
The width of the distribution of $z_1$ and of $\log g$ 
is therefore  narrower
in 3D than in in the Q1D. 
This agrees with numerical data presented in figure \ref{fi3}. Note
also an asymmetry of the distribution: the probability to find 
$\log g=\langle\log g\rangle +\delta$ is higher than
the probability to find 
$\log g=\langle\log g\rangle -\delta$ ($\delta>0$).

\begin{figure}
\epsfig{file=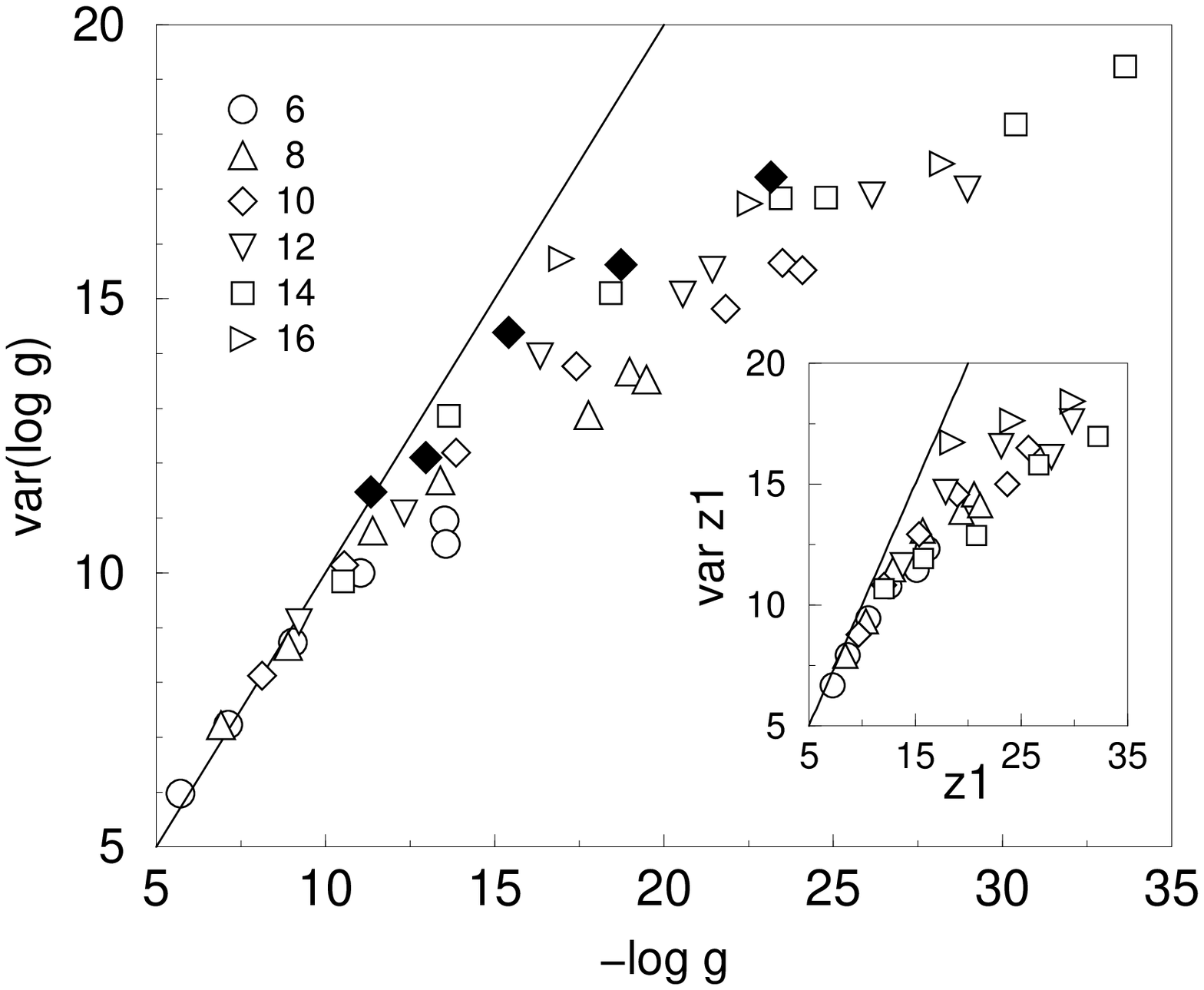,width=8.25cm}
\vspace*{2mm}
\caption{var $\log g$  vs $-\log g$ for various $W$ and 
$L$. Solid line is linear dependence var $\log g = -\log g$. Data
lie on the same curve with exception of some points for $L=6$ and 8,
which represent systems with localization length smaller than 
inter-atomic distance. For completeness, we calculated also var $\log g$
for some systems with non-zero Fermi energy (= 4.4 and 2.5) with $L=12$, full diamonds.
}\label{fi7}
\end{figure}

To support one-parameter scaling by more quantitative arguments we
analyzed also statistical properties of the parameters $z$.
Contrary to weakly disordered Q1D systems,  we
have to take into account also mutual correlations of $z_1$ and higher
$z$ in 3D. It is therefore not clear in advance, whether or not  the one-parameter scaling
theory is valid in this regime.
However, Figure \ref{fi5} shows  that  the differences $\langle z_i-z_1\rangle$ 
are unambiguous functions of $\langle z_1\rangle$. 
Validity of the one-parameter scaling in the localized regime is confirmed also by  the data 
in  figure  \ref{fi4}   which shows that   var $z_1$ is an 
unambiguous  function of  the mean value  
$\langle z_1\rangle$ in 2D even when
disorder is strong.  The same  seems  to hold also  for 3D systems
(figure \ref{fi7})
In figure \ref{fi7} we present also the 
 $\langle \log g\rangle$-dependence 
of var log $g$.  In the last case we find
deviations from scaling only when the localization length $\xi$ becomes 
comparable with inter-atomic distance ($\xi\approx 1$). 
We analyzed also systems with non-zero Fermi energy. This was motivated by
Ref. \cite{slevin}, where separate  var $z$ {\sl vs} $\langle z\rangle$ -  
relations were found in systems  with different  Fermi energy.
Our present data, however,  scale to the same curve as that for the 
zero Fermi energy (Figure \ref{fi7}). However,
we cannot guarantee  that more detailed statistical analysis
will not show small
deviations from the one-parameter scaling, similar to that observed
in one-dimensional systems \cite{slevin}.

\section{Conclusion}

We have shown numerically that the shape of the conductance distribution   
depends on the system dimension when the system leaves  metallic regime.
Qualitative explanation of this difference is based on the known
spectra of the parameters $z$.
In  weakly disordered quasi-one dimensional systems  
the  spectrum of $z$ is linear independently on the
dimension and the system length. However,
in the cubic $d$-dimensional samples 
the  spectrum depends on the 
dimension of the system when the disorder increases. 

We presented and compared the conductance distributions for Q1D and 3D systems
in the intermediate (critical) regime. For both of them we found non-analytical behavior
in $g=1$ and exponential decrease of $P(g)$  for $g>1$. The 
distributions differ only quantitatively.

We have shown  for the first time that the differences $\langle z_2-z_1\rangle$ increase logarithmically with the system size in two-dimensional strongly
disordered systems. 
In three dimensional strongly disordered systems 
the differences  
$\langle z_2-z_1\rangle $ and
$\langle z_3-z_1\rangle $ 
are constant and  independent  on the 
system size and  on the strength of the disorder.  Is agrees with
previous results  \cite{JPCM}.
The contributions to the conductance 
from  the first, the second and higher  channels 
have  therefore  the same system-size dependence $\sim\exp -z_1$.
Although the contribution of higher channels to the mean conductance is small, 
the shape of the spectra of $z$ influences strongly the form of the 
distribution of the $\log g$, which becomes narrower in 3D than in the
weakly disordered quasi-one dimensional systems.
The  one-channel approximation is therefore not sufficient for the entire
 description of the insulating regime in three dimensions. 

Our numerical data 
confirm that $z_2$, $z_3$ together with var $z_1$  are unambiguous 
functions of $z_1$.  We believe therefore, that, in
spite of the non-linearity of the  spectrum of $z$,
the random matrix theory  can be generalized
to the description of the non-metallic systems \cite{chen,JPCM}.
Such a generalization would approve applicability of 
the one-parameter scaling theory  both to the critical and the strongly insulating regime.

\medskip
 
\noindent This work was supported by Slovak Grand Agency VEGA, Grant n. 2/7174/20. Numerical data were partially collected using the computer Origin 2000
in the Computer Center of the Slovak Academy of Sciences.

\newpage

\begin{thebibliography}{99}
\item[*] E-mail address:  markos@savba.sk
\bibitem{AALR} E. Abrahams, P.W. Anderson, D.C. Licciardello and T.V. Ramakrishnan, Phys. Rev. Lett. {\bf 42}, 673 (1979)
\bibitem{MacKK} A. MacKinnon and B. Kramer, Phys. Rev. Lett. {\bf 47}, 1546 (1981)
\bibitem{q1d}
B. Bulka, B. Kramer and A. MacKinnon, Z. Phys. {\bf B60}, 13 (1985);
B. Kramer, K. Broderix, A. MacKinon and M. Schreiber, Physica {\bf A 167}, 163  (1990);
A. MacKinnon, J. Phys.: Condens. Matt. {\bf 6}, 2511 (1994);
K. Slevin and T. Ohtsuki, Phys. Rev. Lett. {\bf 82}, 382 (1999);
P. Cain, R. A. R\"omer and M. Schreiber, Ann. Phys. (Leipzig) {\bf 8}, SI-33 (1999)
\bibitem{shapiro} P.W. Anderson, D.J. Thouless, E. Abrahams and D.S. Fisher, Phys. Rev. {\bf B 22}, 3519 (1980);
B. Shapiro, Phys. Rev. Lett. {\bf 65}, 1510  (1990)
\bibitem{stone} P.A. Lee, A.D. Stone and H. Fukuyama, Phys. Rev. {\bf B 35}, 1039  (1987) 
\bibitem{DMPK} O.N. Dorokhov, JEPT Letter {\bf 36}, 318 (1982); 
P.A. Mello, P.Pereyra and N. Kumar Ann. Phys. (NY) {\bf 181}, 290   (1988)
\bibitem{pichard} J.-L. Pichard, in Quantum Coherence in Mesoscopic Systes,
Ed. by B. Kramer, NATO ASI Ser B {\bf 254} 369 (1991)
\bibitem{beenak} C.W.J. Beenakker and R. Rajaei, Phys. Rev. {\bf B 49}, 7499 (1994) 
\bibitem{epsilon} B.L. Altshuler, V.E, Kravtsov and I.V. Lerner, Soviet JETP {\bf 64}, 1352 (1986); JETP Lett. {\bf 43}, 441 (1986); 
\bibitem{epstwo} B. Shapiro and A. Cohen, Int. J. Mod. Phys. {\bf B6}, 1243 (1992)
\bibitem{MKa} P. Marko\v s and B. Kramer, Phil. Mag. {\bf B68}, 357  (1993) 
\bibitem{ostatni}  N. Giordano, Phys. Rev. {\bf B38}, 4746 (1988) 
\bibitem{ando} P. Marko\v s, Europhys. Lett. {\bf 26}, 431  (1994) 
\bibitem{SO} K. Slevin and T. Ohtsuki, Phys. Rev. Lett. {\bf 78}, 4083 (1997); {\sl ibid} {\bf 82}, 689  (1999);
C.M. Soukoulis, Xiaosha Wang, Qiming Li and M.M. Sigalas,   Phys. Rev. Lett. {\bf 82}, 688  (1999)
\bibitem{PRL} P. Marko\v s, Phys. Rev. Lett. {\bf 83}, 588 (1999)
\bibitem{RMS} M. R\"uhl\"ander, P. Marko\v s and C. Soukoulis, cond-mat/0104394
\bibitem{soukou}  Xiaosha Wang, Quiming Li and    C.M.  Soukoulis, Phys. Rev {\bf B58}, 3576 (1998) 
\bibitem{SMO} K. Slevin, P. Marko\v s and T. Ohtsuki, Phys. Rev. Lett. {\bf 86}, 3594 (2001)
\bibitem{pozn} Providing that system size L is smaller than localization length.
\bibitem{MW} K.A. Muttalib and P. W\"olfle, Phys. Rev. Lett. {\bf 83}, 3013 (1999)
\bibitem{MWa} A. Garc\'{\i}a-Mart\'{\i}n and J.J. S\'aenz, Phys. Rev. Lett. {\bf 87}, 116603 (2001)
\bibitem{MKb} B. Kramer and M. Schreiber, J. non cryst. Solids {\bf 114}, 330 (1989)
\bibitem{MKc} P. Marko\v{s} and B. Kramer, Ann. Physik (Leipzig) {\bf 2}, 339 (1993)
\bibitem{muttalib} K.A. Muttalib, Phys. Rev. Lett. {\bf 65}, 745 (1990)
\bibitem{landauer} R. Landauer, IBM Res. Dev. {\bf 1}, 223 (1957); E.N. Economou and C. Soukoulis, Phys. Rev. Lett. {\bf 46}, 618 (1981); D. S. Fisher and P.A. Lee, Phys. Rev. {\bf B 23}, 6851 (1981); 
\bibitem{aando} T. Ando, Phys. Rev. {\bf B 44}, 8017 (1991);
\bibitem{imry} I. Ymry, Europhys. Lett. {\bf 1}, 249 (1986) 
\bibitem{nato} P. Marko\v s and M. Henneke, J. Phys.: Condens. Matt. {\bf 6}, L765  (1994); P. Marko\v s, NATO ASI Ser. E  {\bf 291}, 99 (1995) 

\bibitem{JPCM} P. Marko\v{s}, J. Phys.: Condens. Matt. {\bf 7}, 8375 (1995)
\bibitem{slevin} K. Slevin and J.B. Pendry: J. Phys.: Condens. Matt. {\bf 2}, 2821 (1990) 
\bibitem{chen} Y. Chen, H.E.M. Ismail and K.A. Muttalib, J. Phys.: Condens. Matt. {\bf 4}, L417 (1992)

\end{thebibliography}
\end{document}